\documentclass[twocolumn,letterpaper, showpacs,preprintnumbers,amsmath,amssymb,pre]{revtex4}
\usepackage{graphicx}% Include figure files
\usepackage{dcolumn}% Align table columns on decimal point
\usepackage{bm}% bold math

\begin{document}

%\preprint{APS/123-QED}

\title{Microrheological Characterisation of Anisotropic Materials}

\author{I A Hasnain}

\author{A M Donald}%
\email{amd3@cam.ac.uk}
\affiliation{
Biological and Soft Systems,
Cavendish Laboratory,
University of Cambridge,
Madingley Road, Cambridge, United Kingdom CB3 0HE
}

\date{\today}

\begin{abstract}
We describe the measurement of anisotropic viscoelastic moduli in complex soft materials, such as biopolymer gels, via video particle tracking microrheology of colloid tracer particles. The use of a correlation tensor to find the axes of maximum anisotropy without prior knowledge, and hence the mechanical director,  is described. The moduli of an aligned DNA gel are reported, as an application of the technique; this may have implications for high DNA concentrations \textit{in vivo}. We also discuss the errors in microrheological measurement, and describe the use of frequency space filtering to improve displacement resolution, and hence probe these typically high modulus materials.
\end{abstract}

%\keywords{microrheology liquid crystal DNA anisotropic errors}%Use showkeys class option if keyword
                              %display desired
\maketitle

Many important biological molecules are capable of forming liquid crystalline phases due to their inherent stiffness and aspect ratio. In this paper we demonstrate how a development of existing microrheological techniques offers a means to characterize local anisotropic properties in such media.

We believe that this technique is unique in its ability to determine the mechanical properties of typical anisotropic materials, where the director may vary over a mesoscopic lengthscale. Our approach also utilises several improvements to existing microrheology measurements, which are likely to be particularly important for such high modulus, anisotropic materials for which there are particular challenges. Although we have chosen to work with DNA, many materials, both naturally occuring and synthetic, form anisotropic phases. Examples of such materials of biological importance include aqueous biopolymer gels, including DNA, but also cellulose, xanthan and F-actin, amongst others. Anisotropy may also occur in intracellular materials, to which standard 2-D microrheology has previously been applied \cite{active_microrheology_cells_lubensky, Living_cells_micro_kuo, Living_cells_micro_wirtz}.

Particle tracking microrheology is a technique that has attracted much interest in experimental soft condensed matter as a method for measuring the viscoelastic properties of soft materials on the micron scale. The mechanical response of such a material may be probed by monitoring the time evolution of the displacement of embedded microscopic tracer particles subject to thermal motion.

The idea of using embedded particles to probe material properties was first described in a seminal paper by Freundlich and Seifriz \cite{Freundlich_seminal}, studying gelatin gels using magnetic beads in an external field. However these experiments suffered from the use of aspherical particles, limited force control and inaccuracy in the detected particle position. The technique was rediscovered by Mason and Weitz \cite{MasonWeitz95} who proposed an extension to the familiar Stokes-Einstein diffusion equation to complex, frequency dependent moduli,
\begin{equation}
\label{GSER}
\tilde{G}(s) = \frac{d k_B T}{3 \pi a s \langle r(s)^2 \rangle}
\end{equation}
where the real viscoelastic shear modulus $\tilde{G}$ of a material at a temperature $T$ may be calculated as a function of Laplace frequency, $s$, for beads of radius $a$, with a displacement $r$, in $d$ dimensions, given the mean squared displacement (MSD), $\langle r^2(\tau) \rangle = \langle \left | r(t+\tau)-r(t) \right |^2 \rangle$. In addition to improved experimental techniques, this presentation of a generalised Stokes-Einstein relation (GSER) has encouraged renewed interest, and resulted in many advances, greatly expanding the power of microrheology as a technique (for a review see e.g. \cite{Micro_high_throughput}). 

From the MSD, under the same assumptions as the generalised Stokes-Einstein relation (GSER) (Eqn.\ref{GSER}) we can calculate \cite{Xu_creep_acta} the creep compliance by multiplication with a constant,
\begin{equation}
\label{creepcompliance_eqn}
\Gamma(\tau) = \frac{\pi a}{k_B T} \langle r^2(\tau) \rangle
\end{equation}
or extract the frequency dependent complex viscoelastic shear modulus via a Laplace or Fourier transform \cite{Mason_algeb_LT_rheolacta, Mason_algebraic_LT, Gittes_Schmidt}. If the material is a gel, i.e. a material composed of a polymer network in a fluid medium,  then the network spacing (mesh size) may also be determined via particle tracking \cite{Val_pore_size}.
\begin{figure}
\includegraphics[width=3in]{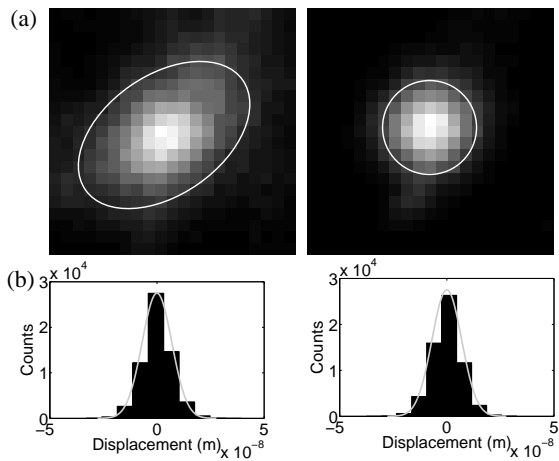}
\caption{\label{particlewalks_fig} (a) Measured probability density plot of probe positions in the aligned DNA gel described later (left) and a similar unsheared DNA sample (right), as an example of an isotropic material. The plots are calculated by collapsing probe tracks (longer than 50 steps) over all available time to a common origin, and counting the occurrences in discrete bins. The angle of the surrounding ellipse is set at 35$^\circ$ from the horizontal, which is the value calculated from the eigenvector analysis for the shortest lag time later. A good correspondence is seen. (b) Displacement distribution in x (left) and y (right) for $\tau$ = 1/25s. Visually the diffusion is Gaussian, and this may be verified by calculating the excess kurtosis of the distributions (0.8 in the $x$-direction and 0.7 in the $y$-direction). Thus the degree of heterogeneity is relatively small, and may be neglected.}
\end{figure}
% End intro

% Background?
Conventional microrheology has so far used the 2-dimensional MSD in Eq. (\ref{GSER}), but we choose to decompose the displacement into 2 orthogonal 1-dimensional displacements. Although an individual random walk is spatially anisotropic \cite{Aspherical_rand_walk_Gasperi}, the probability density function averaged over a number of walks will be circularly symmetric in an isotropic material. In a mechanically anisotropic material, the ensemble average of the random walks will also be elliptically symmetric (Fig. \ref{particlewalks_fig}) with major and minor axes along the mechanical director. The GSER thus allows us to calculate the dissipative and elastic properties of the gel along specific axes, subject to the same conditions as the Stokes derivation: no-slip boundary conditions on a sphere in an incompressible continuum fluid, with no inertia. In addition the implicit assumption is made that the Stokes drag for viscous fluids may be generalised to viscoelastic materials at all frequencies, and that the probe particles have a negligible effect on the material \cite{Mason_algeb_LT_rheolacta, Levine_Lub_limitations}. Although in general viscosity is an 81 element 4th rank tensor, many of these elements are either zero or repeated in typical cases. For example a nematic liquid crystal has 5 independent intrinsic viscosities describing diffusion. Our case is is analogous to the drag on a sphere moving though a nematic liquid crystal \cite{Ruhwandal_nematicdrag, Stark_nematicdrag, Loudet} and so, since we have a rotational symmetry axis parallel to the director, the Brownian motion is governed by two independent diffusion coefficients where $D_{\perp / \parallel} = k_b T / (6\pi \eta_{\perp / \parallel}) a$, as the derivation of the Stokes-Einstein relation is separable in spatial dimension. Generalising to complex viscosity, we thus expect the GSER to hold in anisotropic materials, with two independent complex viscosities $\eta^{*}_{\perp}$ and $\eta^{*}_{\parallel}$. The ratio, $\eta^{*}_{\perp}/\eta^{*}_{\parallel}$ is therefore a measure of the anisotropy of the system. Similarly to the isotropic case, it should be noted that there is no rigorous theoretical backing for the extension to complex moduli, only phenomeological justification.

The probe particles themselves may be tracked via a number of methods. We choose real-space multiparticle `single-point' video tracking due to both the large amount of information available for analysis, and the relative ease of setup. Video is acquired with a Zeiss-Axioplan optical microscope with a 100x oil immersion lens (N.A. = 1.30) and conventional commercial CCD video camera (Sony SCC-DC138P PAL), directly into a computer using a Hauppauge Impact VCB capture card. The analysis is performed using custom-written MATLAB scripts. To detect the small Brownian displacements of colloidal particles in high modulus materials we need as high a spatial resolution as possible. This may be measured by determining the apparent displacement of fixed spheres, following \cite{StatDynErrors}. We use a 2-dimensional bandpass filter to select appropriate spatial frequencies in Fourier space, and thus filter out noise and background variation. Regions above an intensity threshold are selected, then checks for eccentricity and shape solidity are performed to determine the quality of the probe images. Finally the centroid of the region is calculated, weighted by the original pixel intensity. By this method we have achieved a 1D displacement resolution of up to 3nm, with 4nm in typical operation.

\begin{figure}
\includegraphics[width=2in]{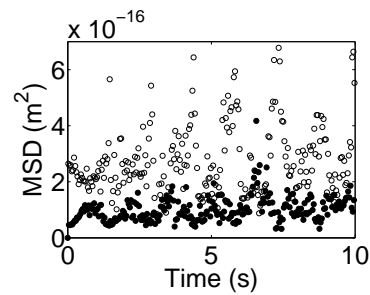}
\caption{\label{deinterlace_error_fig}Mean squared displacement of 0.497$\mu$m polystyrene latex colloidal spheres dried on a glass slide, calculated from all independent lag times. The deinterlaced video fields appear displaced from each other by $\sim$ 0.15 pixels (most clearly seen at short lag times), varying by $\sim$ 0.06 pixels with time (for clarity even lag times, in frames, are denoted by empty symbols). This represents a significant error for high modulus materials.}
\end{figure}

Since the video data in conventional cameras is recorded in an interlaced manner, in which the two fields of alternate lines are recorded either 1/50 or 1/60 of a second apart in each frame, one typically de-interlaces the fields and analyses each field separately. However, by tracking fixed colloidal spheres at high resolution, we show in Fig. \ref{deinterlace_error_fig} that the two fields are displaced from each other in a time varying fashion. The source of this displacement is unclear but it means that de-interlacing may be inadvisable when tracking small displacements, and is not carried out in the current work.

High modulus gels often have some in-built shear stress when they are made, which relaxes over a very long time period. In addition one may wish to investigate the microrheological properties of sheared, flowing, or even living samples. However, probe particles in such a system will be subject to a net flow field, in addition to Brownian motion. Provided that the flow field is uniform, $\langle \Delta r(\tau) \rangle = 0$ for a random walk, so the net drift velocity can be calculated from the ensemble average and then subtracted. If the flow field varies across the field of view this approach is clearly invalid, and if the velocity field is a function of depth the situation is complicated \cite{zumofen_enhanced_diffusion}, but in practice we find this simple approach to be adequate in this work, where we measure the drift due to relaxation as $\sim$ $3\mu$m/min.

\begin{figure}
\includegraphics[width=2in]{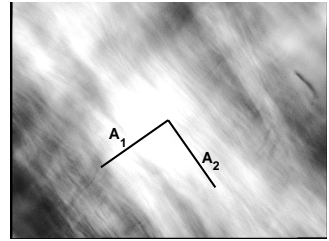}
\caption{\label{fig_eigenpicture}The sheared 11mg/ml DNA sample at 298K under horizontal-vertical crossed polarizers. Alignment is relatively weak. Light areas show alignment, with the DNA chain backbones lying approximately 45$^\circ$ counterclockwise from the vertical (checked via insertion of a quarter waveplate at 45$^\circ$ to the polarizers). The calculated eigenvectors are superimposed in black, and show the mechanical alignment corresponds well with the optical birefringence, as expected. The calculated angle is 35$^\circ$ counterclockwise.} 
\end{figure}

We write the displacement in tensor form to remove coordinate system dependence from the arbitrary x (horizontal) y (vertical) Cartesian axes with the origin in the top-left in the video. The eigenvector relation (Eqn. \ref{eigen_eqn}) can then be written, and the resulting eigenvectors give the basis in which displacements along the two orthogonal axes are least correlated, and hence maximally anisotropic. We use the eigenvectors calculated from the shortest lag time, $\tau$, for which there are the most events and hence the smallest error, to find the new basis and hence recalculate displacements in these coordinates. For an anisotropic material one would expect the new axes to lie parallel and perpendicular to the average optical director. By solving the equation for the specific example shown in Fig. \ref{fig_eigenpicture}, and determining the eigenvectors (superimposed on Fig. \ref{fig_eigenpicture}), we can indeed see the correlation of the optical alignment with the directions of anisotropy. The directionality and strength of alignment in the material can thus be determined from particle tracking alone, with no prior knowledge required. We will discuss the nature of the sample below.
\begin{equation}
\label{eigen_eqn}
\begin{bmatrix}
\langle (\Delta x(\tau))^2 \rangle     &   \langle \Delta x(\tau) \Delta y(\tau) \rangle \\
\langle \Delta x(\tau) \Delta y(\tau) \rangle    & \langle (\Delta y(\tau))^2 \rangle
\end{bmatrix} \mathbf{A} = \lambda \mathbf{A}
\end{equation}

The MSD must then be computed. This is a concise representation of the rheological data, related to the creep compliance by a constant (Eqn. \ref{creepcompliance_eqn}). We need to determine the errors in microrheological data, with a particular view to judging whether anisotropy is significant, so the following expression is derived, making the assumption that a `typical' value for $x$ is $\sqrt{\langle\Delta x^2 (\tau)\rangle}$. Assuming Gaussian statistics, we have that $\sigma_{x^2} = 2 \sigma_x x$ and so,
\begin{eqnarray}
\label{MSD_error_eqn}
\sigma_{\langle\Delta x^2 (\tau)\rangle} & = & \frac{2\sigma_x x}{\sqrt{N}} \sim \frac{2 \sigma_x \sqrt{\langle\Delta x^2 (\tau)\rangle}}{\sqrt{N}}  \nonumber \\ 
 & \sim & \frac{2 \langle\Delta x^2 (\tau)\rangle}{\sqrt{N}}
\end{eqnarray}
where $N$ is the number of events contributing to $\langle\Delta x^2 (\tau)\rangle$. The expression is of the appropriate form as $\sigma_x = \sqrt{\langle\Delta x^2 (\tau)\rangle}$ in this case. In addition, the noise at the resolution limit results in a `static error' \cite{StatDynErrors} which is exhibited as a constant offset added to $\langle\Delta r^2 (\tau)\rangle$, and thus can be subtracted.

A number of methods have been proposed to arrive at a complex shear modulus \cite{Gittes_Schmidt, Dasgupta_LT, Mason_algeb_LT_rheolacta, MasonWeitz95}. We use a simple discrete Laplace transform,
\begin{equation}
\label{DLT_eqn}
\tilde{x}(s) = \frac{\tau_N}{N} \sum^{N}_{n=0} x(\tau_n)e^{-s\tau_n}
\end{equation}
where $N$ is the number of samples spread over the interval $\tau=0$ to $\tau=\tau_N$, followed by 1st order polynomial spline fitting and analytic continuation into the complex plane, substituting $i\omega$ for $s$ in the locally fitted form \cite{MasonWeitz95, Bird_analycont}. No assumptions are made about the form of the response which is well approximated by a linear spline fit over a 5 point window, and truncation errors are found to be small for a discrete Laplace transform given a typical range of data in time (1/25s to 60s).

To demonstrate the utility of microrheology as a tool for probing anisotropic materials, we apply the method described above to shear-aligned deoxyribonucleic acid (DNA). DNA in water, with some ammonium acetate as a counter ion, forms well-documented liquid crystalline phases \cite{DNA_liq_cryst} at high concentrations. We look at genomic DNA (Fluka, Deoxyribonucleic acid sodium salt from herring testes. Product: 31162) at a moderate concentration (11mg/ml) at 298K in 0.25M ammonium acetate, contained in a cell made of a metal washer (0.80mm deep) sandwiched by a glass microscope slide and coverslip, holding 12$\mu$l of sample. The sample is left to relax for 1 hour, then shear stress is applied by moving the coverslip, to induce alignment, which may be monitored by optical birefringence (Fig. \ref{fig_eigenpicture}). We use 0.3$\mu$m diameter probes (polystyrene latex, Agar Scientific) which are much larger than the expected mesh size of 15nm (calculated assuming uniform volume filling), with our field of view set well away from the cell surfaces. The video is kept interlaced, as discussed previously, and 120 seconds is analyzed.
\begin{figure}
\includegraphics[width=2.8in]{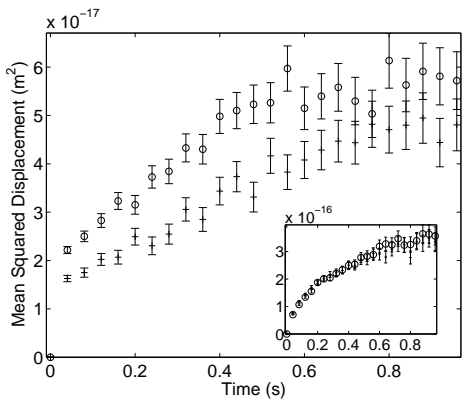}
\caption{\label{fig_dna_msd}1-dimensional MSDs along calculated eigenvectors for sheared 11mg/ml DNA at 298Kß. Crosses (+) correspond to displacements along the axis approximately parallel to the optical director, and circles ($\circ$) to displacements perpendicular to this.  The static resolution error is taken as 4nm, and subtracted. The MSD is proportional to the creep compliance \cite{Xu_creep_acta}. Inset: MSD for the same DNA at 4.9 mg/ml, where there was no visible birefringence. There is now no visible anisotropy in the MSD. Note that the shape of the response is similar to the anisotropic case - this suggests that the GSER should hold in anisotropic materials.}
\end{figure}

\begin{figure}
\includegraphics[width=3in]{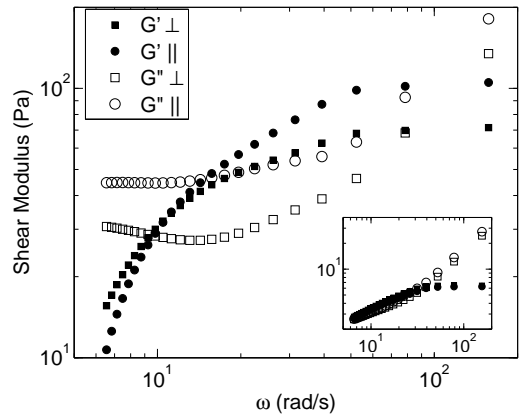}
\caption{\label{fig_dna_moduli}1-dimensional elastic modulus, $G^{\prime}(\omega)$ and loss modulus, $G^{\prime\prime}(\omega)$, along calculated eigenvectors for sheared 11mg/ml DNA at 298K. Filled circles correspond to $G^{\prime}$ along the axis approximately parallel to the optical director, and filled squares to $G^{\prime}$ perpendicular to this. Open symbols refer to $G^{\prime\prime}(\omega)$ in these directions. The static resolution error is taken as 4nm, and subtracted from the MSD before calculation. Inset: Moduli of the same DNA at 4.9 mg/ml (approximately the mean concentration of DNA in a human cell \cite{Wirtz_DNA}). The sample is isotropic in its mechanical response, and the shape is comparable to the anisotropic case, suggesting the validity of the GSER in anisotropic materials.}
\end{figure}

Although the sample exhibited only weak optical alignment, we find significant mechanical differences between the axes parallel and perpendicular to the optical director. Figure \ref{fig_dna_msd} plots the MSD, showing greater motion perpendicular to the optical director than parallel. This seems reasonable in a network of stretched chains, with the optical director lying along the chain backbone. The viscoelastic moduli are also calculated (Fig. \ref{fig_dna_moduli}). The material is clearly stiffer to shear deformation in the direction along which the chains are, on average, already stretched. This is as we would expect for a polymer network which has been stretched uniaxially, where a proportion of chains are close to fully extended between entanglements/ crosslinks in this direction. It is interesting to note that there appears to be a cross-over in the elastic modulus, where the stretched direction has a lower elastic modulus than the perpendicular direction at frequencies below 3Hz, the explanation for which is unclear. The plateau in $G^{\prime}$ is typical of a gel in the elastic region of its frequency response. Some studies \cite{Wirtz_DNA,Chen_DNA} have been done on DNA at lower concentrations, which indicate precursors to this elastic regime.

%Summary
We have presented a method to look at the viscoelastic properties of anisotropic materials on typical (mesoscopic) lengthscales. The technique both permits the identification of the principal axes of the system, and the associated mechanical anisotropy to be characterized as a function of frequency. Some rheological data for mechanically aligned DNA is presented. These observations may have implications for high DNA concentrations \textit{in vivo}. Several improvements to microrheology measurements have been described, with a particular focus on high modulus anisotropic materials.
% Future
Many intracellular materials appear anisotropic, and the approach presented allows the elucidation of the mechanical properties of the materials, potentially \textit{in vivo} \cite{Wirtz_nucleus}. Higher modulus materials should be able to be explored using fluorescent probes with diameters in the 10-100nm range. It should be possible to extend the technique to `2-point' \cite{2particle_Weitz} cross-correlation techniques to ensure that probe effects on the network \cite{Colloid_surf_chem_Kuo, Colloid_surf_chem_PT} can be neglected \cite{2particle_actin_Weitz, Levine_Lub_limitations}. We anticipate that the technique described will also be of use in determining the anisotropic pore sizes of materials which form an anisotropic network.
\begin{acknowledgments}
We would like to thank Pietro Cicuta, Tim Hosey, Heather Houghton, Mark Krebs and Salman Rogers for valuable discussions. This work was supported by BBSRC. 
\end{acknowledgments}

\appendix

\bibliography{aniso1bib1}

 \end{document}